\documentclass[letterpaper]{IEEEtran}
\usepackage{color}
\usepackage{amsmath,amsfonts}
\usepackage{algorithmic}
\usepackage{array}
\usepackage{bigstrut}
\usepackage{multirow}

\usepackage[caption=false,font=normalsize,labelfont=sf,textfont=sf]{subfig}
\usepackage{booktabs}
\usepackage{bm}
\usepackage{textcomp}
\usepackage{stfloats}
\usepackage{url}
\usepackage{verbatim}
\usepackage{graphicx}

\usepackage{cite}
\def\BibTeX{{\rm B\kern-.05em{\sc i\kern-.025em b}\kern-.08em
		T\kern-.1667em\lower.7ex\hbox{E}\kern-.125emX}}
\usepackage{balance}
\usepackage{multirow}

\UseRawInputEncoding

\usepackage{amsmath}
\usepackage{amssymb}

\usepackage{graphicx}
\usepackage{float}
\usepackage{subfig}
\begin{document}
\title{Deep Learning Based Joint Channel Estimation and Positioning for Sparse XL-MIMO OFDM Systems
}
\author{Zhongnian Li, Chao Zheng, Jian Xiao, Ji Wang, \textit{Senior Member, IEEE}, Gongpu Wang, Ming Zeng, 	
					
	and Octavia A. Dobre, \textit{Fellow, IEEE}
	\thanks{Zhongnian Li, Chao Zheng, Jian Xiao and Ji Wang are with the Department of Electronics and Information Engineering, College of Physical Science and Technology, Central China Normal University, WuHan 430079, China (email: zhongnian.li@ccnu.edu.cn; cccjh@mails.ccnu.edu.cn; jianx@mails.ccnu.edu.cn;  email: jiwang@ccnu.edu.cn). (\textit{Corresponding author: Ji Wang})}
	\thanks{Gongpu Wang is with the Beijing Key Laboratory of Transportation Data Analysis and Mining, Beijing Jiaotong University, Beijing 100044, China (email: gpwang@bjtu.edu.cn).}
	\thanks{Ming Zeng is with the Department of Electric and Computer Engineering, Laval University, Quebec City, Canada (email: ming.zeng@gel.ulaval.ca).}
	
	\thanks{Octavia A. Dobre is with the Faculty of Engineering and Applied Science, Memorial University, St. John's, NL A1C 5S7, Canada (e-mail: odobre@mun.ca).}

	}
\maketitle
\begin{abstract}
This paper investigates joint channel estimation and positioning in near-field sparse extra-large multiple-input multiple-output (XL-MIMO) orthogonal frequency division multiplexing (OFDM) systems. To achieve cooperative gains between channel estimation and positioning, we propose a deep learning-based two-stage framework comprising positioning and channel estimation. In the positioning stage, the user's coordinates are predicted and utilized in the channel estimation stage, thereby enhancing the accuracy of channel estimation. Within this framework, we propose a U-shaped Mamba architecture for channel estimation and positioning, termed as CP-Mamba. This network integrates the strengths of the Mamba model with the structural advantages of U-shaped convolutional networks, enabling effective capture of local spatial features and long-range temporal dependencies of the channel. Numerical simulation results demonstrate that the proposed two-stage approach with CP-Mamba architecture outperforms existing baseline methods. Moreover, sparse arrays (SA) exhibit significantly superior performance in both channel estimation and positioning accuracy compared to conventional compact arrays.
\end{abstract}
\section{Introduction}
As sixth-generation integrated sensing and communication (ISAC) technology aims to advance sensing and communication capabilities while maximizing energy efficiencies, multiple-input multiple-output (MIMO) systems must evolve in spatial resolution to enable both enhanced spectral efficiency and precise sensing of densely located targets~\cite{Li2024}. One promising solution is extremely large-scale MIMO (XL-MIMO), which extends traditional massive MIMO by deploying a larger number of antennas~\cite{Wang2023XL_MIMO}. However, conventional compact arrays (CAs) with half-wavelength element spacing face prohibitively high hardware costs~\cite{Wang2018Analysis}. A promising solution to this issue is sparse MIMO, an alternative MIMO architecture that achieves larger aperture without increasing the number of array elements~\cite{Xu2025Integrated}. In addition, sparse XL-MIMO systems with element spacing exceeding half-wavelength enable larger element separations, which significantly enhance spatial resolution. Accurate channel estimation and positioning are crucial for sparse XL-MIMO systems due to limited spatial diversity, which affects signal resolution. 

Up to now, existing research on sparse XL-MIMO is still in early stages. The authors in~\cite{Chen2025NearFieldSparseUPA} derived closed-form expressions for the power distribution and main-lobe dimensions of sparse uniform planar arrays around near-field focal points, and proposed a method for estimating the effective degrees of freedom, thereby demonstrating the potential of sparse arrays (SAs) to suppress interference and enhance capacity in XL-MIMO systems. In~\cite{Zheng2020TwoDimensionalDOA}, a tensor-decomposition direction of arrival (DOA) estimator for coprime arrays is devised. By exploiting higher-order singular value decomposition and the array’s inherent low-rank structure, the method interpolates missing sensors and suppresses noise, thus offering enhanced robustness in multi-target, low-signal-to-noise ratio (SNR) environments. In~\cite{6289992}, a direct-multiple signal classification method for nested and coprime arrays is introduced, avoiding coarray ambiguities and improving resolution. The authors in~\cite{8081236} proposed thinned coprime arrays that maintain the virtual aperture and DOA performance while cutting sensor count and cost. The authors in~\cite{Eisenbeis2021SparseArray} designed a SA-aided channel-estimation scheme for subarray-based hybrid beamforming architectures.

Although these studies have advanced channel estimation and positioning in sparse XL-MIMO systems, they adopt a decoupled design paradigm. Under near-field spherical-wave propagation, user positioning and channel estimation are inherently coupled. A separated approach not only degrades accuracy but also demands independent pilot designs and algorithms, resulting in increased resource consumption and complexity drawbacks that are especially detrimental to resource-constrained XL-MIMO systems. Based on our review, a unified approach that concurrently addresses positioning and channel estimation  with SAs in near-field XL-MIMO systems is still missing. Traditional methods encounter performance bottlenecks in high-dimensional, nonlinear, and dynamic channel environments. Deep learning (DL) methods, with their robust feature learning capabilities, can effectively learn complex nonlinear mappings, adapting to these conditions. To address this gap, a novel two-stage DL framework is proposed for sparse XL-MIMO orthogonal frequency division multiplexing (OFDM) systems. The main contributions in this paper are outlined as follows:
	\begin{itemize}
		\item We propose a novel two-stage joint positioning and channel estimation method for sparse XL-MIMO OFDM systems. First, we estimate the user equipment (UE) coordinates, which are then used to calculate the line-of-sight (LoS) channel. Subsequently, the LoS channel and pilot signals are employed to estimate the non-LoS (NLoS) channel, thereby reconstructing the complete channel. This two-stage method not only improves positioning and channel estimation accuracy but also greatly reduces computational complexity.
		\item To address the reduction in spatial correlation caused by the enlarged element spacing in SAs, and inspired by U-Net's capability for local, multi-scale feature extraction together with Mamba's strength in modelling long-range dependencies, we propose the CP-Mamba architecture for joint channel estimation and user positioning in sparse XL-MIMO OFDM systems. In this design, U-Net provides a high-resolution, gradient-stable backbone that complements Mamba global sequence modelling.
		\item Numerical results show that the proposed CP-Mamba network surpasses the latest positioning and channel-estimation techniques. SAs exhibit superior performance in positioning and channel estimation than CAs.
	\end{itemize}
\section{System Model}
 \begin{figure}[!t]
	\centering
	\includegraphics[width=3.5in]{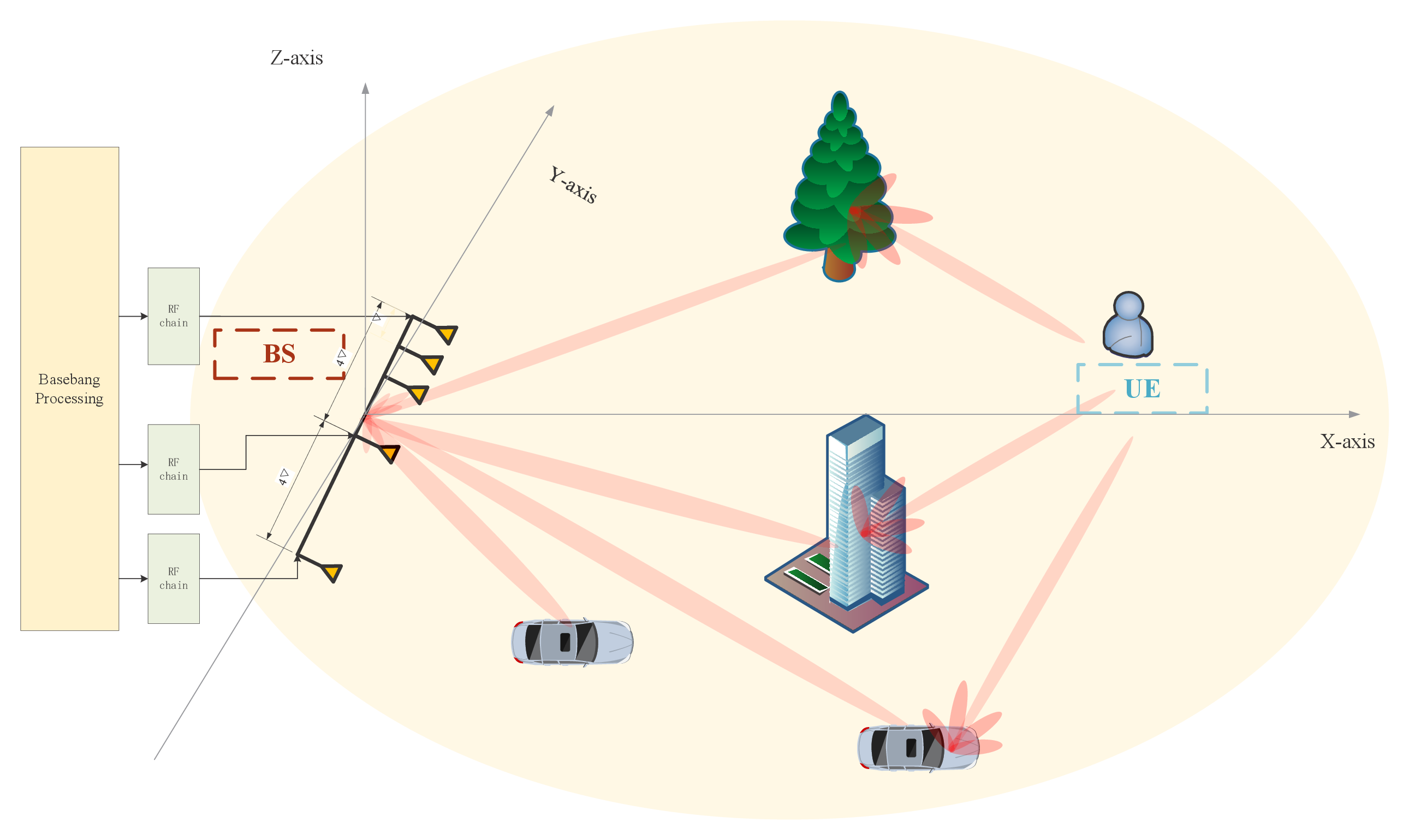}
	
	\caption{{Sparse XL-MIMO system model}.}
	
	\label{Fig1}
	\vspace{-10pt}
\end{figure}
	As shown in Fig.~\ref{Fig1}, we consider a sparse XL-MIMO OFDM scenario in which a base station (BS) with an $N$-element SA serves a single-antenna UE. The BS array center is placed at the global origin. The BS is equipped with only $N_\mathrm{RF}$ radio-frequency (RF) chains ($N_\mathrm{RF} < N$), and assigns $K$ OFDM subcarriers to the UE. By assuming there are a total of $K$ orthogonal subcarriers, the frequency of the $k$-th subcarrier frequency is given by \(
	f_k = f_c - \frac{B}{2} + \frac{(k - 1)\,B}{K}\), where $f_c$ and $B$ denote the carrier frequency and bandwidth, and $\lambda$ is the carrier wavelength.
	We assume that the UE is randomly located in the front half-plane of the BS, with the polar radius following $r$ and denote the azimuth angle of the user relative to the BS as $\theta$. The UE coordinate is $
	\bigl(x_{\mathrm{UE}},\,y_{\mathrm{UE}},\,z_{\mathrm{UE}}\bigr)
	\;=\;\bigl(r\cos\theta,\; r\sin\theta,\;0\bigr).
	$

\vspace{-5pt}	
\subsection{Array Model}
\begin{figure}[!t]
	\centering
	\captionsetup{skip=-4pt}
	\includegraphics[width=2in]{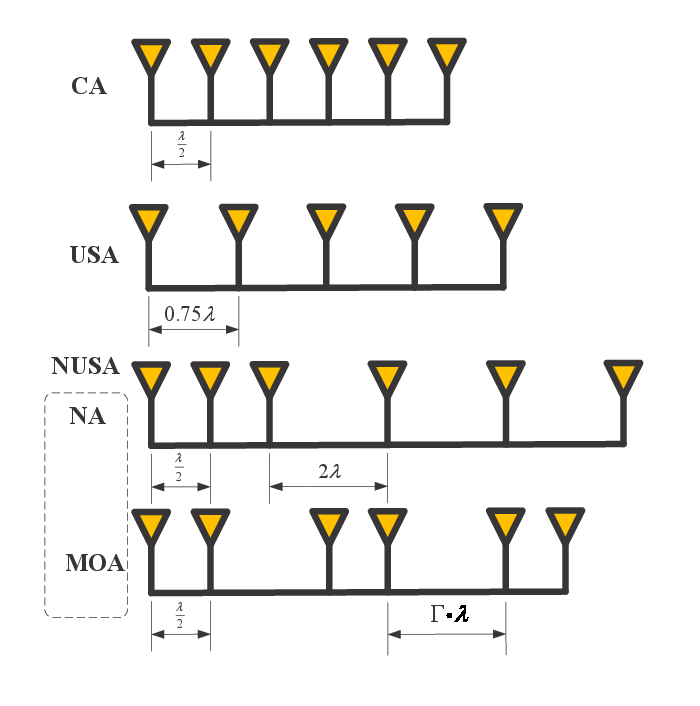}
	\caption{Different array architectures.}
	\label{Fig2}
	\vspace{-10pt}
\end{figure}

As shown in Fig.~\ref{Fig2}, we present the array architectures for CA, uniform-SA (USA) and non-USA (NUSA), which consist of both the modular array (MOA) and the nested array (NA).

\textbf{(1) CA:} Conventional CA employs a uniform linear arrangement of antenna elements, each separated by half the operating wavelength \(d = \frac{\lambda}{2}\). Consequently, the total physical aperture of a CA with \(N\) elements is \(
D = (N - 1)\,d.
\)

\textbf{(2) USA:}  All elements in the USA are uniformly spaced by $d_{\mathrm{USA}} = \eta_{\mathrm{USA}} d$, with sparsity parameter $\eta_{\mathrm{USA}} \geq 1$~\cite{Wang2023Sparse}. Note that the CA is a special case of the USA when $\eta_{\mathrm{USA}} = 1$.

\textbf{(3) MOA:} The MOA comprises $N_1$ modules, each containing $M_1$ antenna elements, totaling $N_1 M_1$ elements. Within each module, the inter-element spacing is $d$, resulting in a module size of $S = (M_1 - 1)d$. Defining the inter-module spacing between module centers as $\Gamma d$ ($\Gamma \geq M_1$), the total aperture is given by $
D = \left[(N_1 - 1)\Gamma + (M_1 - 1)\right] d
$~\cite{Li2024MultiUserModularXL}.

\textbf{(4) NA:} The two-level NA combines an inner uniform linear subarray of $N_1$ elements with spacing \(d\), and an outer sparse subarray of $N_2$ elements with spacing $d_{\text{out}} = (N_1 + 1)d$. This hierarchical configuration ensures enhanced spatial resolution while maintaining structural coherence. The physical sizes of the inner and outer arrays are given by $
S_{\text{inner}} = \{ m d \,|\, m = 1, 2, \dots, N_1 \},$ and $
S_{\text{outer}} = \{ n (N_1 + 1) d \,|\, n = 1, 2, \dots, N_2 \},
$
respectively, while the total aperture is $
D = \left[ N_2(N_1 + 1) - 1 \right] d
$~\cite{Pal2010NestedArrays}.

Due to the SA spacing exceeding $\frac{\lambda}{2}$, grating lobes and angle ambiguity effects arise. In scenarios prioritizing communication reliability and beam control, CA is typically a better choice for channel estimation task. However, the larger spacing enhances angular resolution, making SA more suitable for positioning task. In ISAC scenarios, SA offers significant advantages over CA in hardware cost, mutual coupling control, and spatial reuse, making it more attractive.

\vspace{-10pt}	
\subsection{Channel Model}	
As the SA spacing increases, the Rayleigh distance and near-field region expand. In such scenarios, the near-field channel characteristics are influenced by the angle of arrival (AoA) and distance~\cite{10533725}. To comprehensively model near-field propagation in extremely large-scale array systems, a near-field channel model is adopted.
Suppose a two-level NA or a MOA, to serve a single-antenna UE whose planar positions are known in polar form \((R,\phi)\). In the NA configuration, the array element \((m,n)\) sits at \(x_{m,n}=(N_1+1)\,n\,d\ + m\,d\), for \(n=0,\dots,N-1\) and \(m=0,\dots,M-1\). In the MOA configuration, the array element \((m,n)\) sits at  
\(x_{m,b}=n\,\Gamma\,d + m\,d\), for \(n=0,\dots,N-1\) and \(m=0,\dots,M-1\).

Due to the larger aperture of the SA, the Rayleigh distance increases, causing many UE element links to fall within the electromagnetic near-field. We adopt a nonuniform spherical-wave model, The distance from antenna \(q\) at coordinate \(x_q\) to the $g$-th scatter in the $l$-th cluster is given by\begin{equation}
	r_{l,g,q}
	=\sqrt{R_{l,g}^2 + x_q^2 - 2\,R_{l,g}\,x_q\cos\phi_{l,g}},
	\label{eq:r}
\end{equation}
where $R_{l,g}$ denotes the UE distance from the array center associated with the $g$-th scatter in the $l$-th cluster, $c$ denotes the speed of light, and the corresponding element response is given by
\begin{equation}
	\mathbf{a}_{l,g,q}[k]
	=
	e^{-j\frac{2\pi c}{f_k}\,r_{l,g,q}}.
\end{equation}
 
For the $k$-th subcarrier, the channel $\mathbf{h}[k]$ $\in\mathbb{C}^N$ between the BS and the UE consists of a LoS component $\mathbf{h}_{\mathrm{LoS}}$ and NLoS component $\mathbf{h}_{\mathrm{NLoS}}$, being expressed as 
\begin{equation}
	\begin{split}
		\mathbf{h}[k]
		&= \mathbf{h}_{\mathrm{LoS}} + \mathbf{h}_{\mathrm{NLoS}},\\
		&=  \alpha_{0}[k]\,\mathbf{a}_{0}[k]
		+ \sum_{l=1}^{L}\sum_{g=1}^{G_l} \alpha_{l,g}[k]\,\mathbf{a}_{l,g}[k],
	\end{split}
	\label{eq:h}
\end{equation}
where $L$ is the number of clusters between the UE and BS, $G_{l}$ denotes the number of scatters in the $l$-th cluster, $\mathbf{a}_{l,g}$ denotes the near-field steering vector, and  \(\phi_{l,g}\) denotes the AoA of the $g$-th scatter in the $l$-th cluster between the BS array reference point and the UE.~\cite{Li2023THzXL_RIS}. Therefore, we only keep the LoS term with \(l = 0\), and the LoS channel gain is
$\alpha_{0}[k] = \frac{1}{4 \pi R} \cdot 10^{\frac{Q \cdot R}{10}}$~\cite{11018390}, $R$ denotes the distance from the UE to the BS antenna and  $Q$ is the path loss factor. $\alpha_{l,g}[k] \sim \mathcal{CN}(0, 1)$ denotes the small-scale complex channel gain of the NLoS path~\cite{Li2023THzXL_RIS}.

\vspace{-5pt}
\subsection{Problem Formulation}	
The received uplink pilot signal $\mathbf{y}[p,k] \in \mathbb{C}^{N_{\text{RF}}}$ on the $k$-th subcarrier during the $p$-th time slot, is given by
\begin{equation}
	\mathbf{y}[p,k] = \mathbf{W}[p] \mathbf{h}[k] x[p,k] + \mathbf{n}[p,k],
\end{equation}
where $x[p,k]$ is the pilot symbol transmitted by the UE and $\mathbf{W}[p] \in \mathbb{C}^{N_{\text{RF}} \times N}$ is the analog combiner at the BS~\cite{Li2023THzXL_RIS}. Each element of $\mathbf{W}[p]$ satisfies $
|W_{i,j}[p]|=\frac{1}{\sqrt{N_{\text{RF}}}}$, where $\forall i, j, p$ follows $\mathcal{U}(0,1).$
The noise term is given by $\mathbf{n}[p,k] = \mathbf{W}[p] \bar{\mathbf{n}}[p,k]$, where $\bar{\mathbf{n}}[p,k] \in \mathbb{C}^{N}$ is the original complex Gaussian noise vector, which follows $
\mathcal{CN}(\mathbf{0}_N, \sigma^2 \mathbf{I}_N).
$

Assuming $x[p,k] = 1$, we stack the received signals from all $P$ pilot time slots as $
\mathbf{Y}[k] = 
\begin{bmatrix} 
	\mathbf{y}[1, k]^T \dots \mathbf{y}[P, k]^T 
\end{bmatrix}^T
\in \mathbb{C}^{P N_{\text{RF}}}$ . Define the stacked combiner matrix as $
\bar{\mathbf{W}} = 
\begin{bmatrix} 
	\mathbf{W}[1]^T \dots \mathbf{W}[P]^T 
\end{bmatrix}^T
\in \mathbb{C}^{P N_{\text{RF}} \times N},
$ and the stacked noise vector as $
\mathbf{N}[k] = 
\begin{bmatrix} 
	\mathbf{n}[1, k]^T \dots  \mathbf{n}[P, k]^T 
\end{bmatrix}^T
\in \mathbb{C}^{PN_{\text{RF}}}$~\cite{Li2023THzXL_RIS}.
 Thus, the received signal is expressed as
\begin{equation}
	\mathbf{Y}[k] = \bar{\mathbf{W}} \mathbf{h}[k] + \mathbf{N}[k].
	\label{eq:y}
\end{equation}
\section{Propsed joint channel estimation
	and positioning method}
 We introduce a two-stage method to reduce channel estimation errors caused by antenna sparsity. Considering the reduction in the spatial correlation caused by the enlarged element spacing in SAs, we propose a CP-Mamba architecture by exploiting the U-Net's strengths in local feature extraction and Mamba's ability to capturing global dependencies.
	\begin{figure}[!t]
	\centering
	\captionsetup{skip=-1pt}
	\includegraphics[width=2.5in]{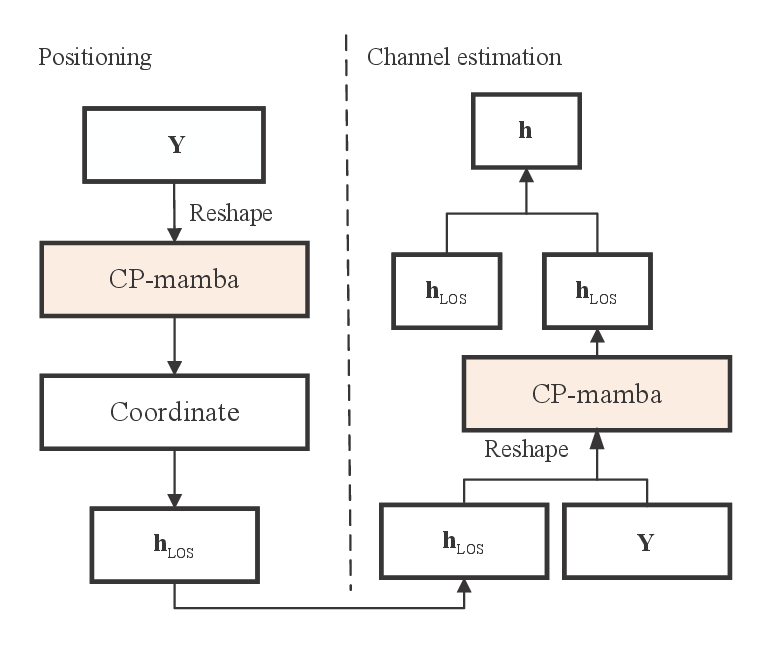}
	\caption{Two-Stage framework for channel estimation and positioning.}
	\vspace{-7.5pt}
	\label{Fig3}
\end{figure}
\vspace{-10pt}
\subsection{The Framework}
 As illustrated in Fig.~\ref{Fig3}, to achieve a synergistic improvement in positioning accuracy and channel estimation quality, the proposed two-stage cooperative procesing framework is illustrated in Fig. 3. The core idea is to leverage wireless signals for precise UE positioning and then use this positional information to optimize channel estimation. Specifically, in the positioning stage, the CP-Mamba network extracts features highly correlated with the UE's spatial coordinates from echo signals transmitted by the BS and reflected by the UE, thereby obtaining the estimated UE location \(\mathbf{\tilde{C}}\), through~\eqref{eq:r}. Based on this estimated positional information, the LoS channel \(\mathbf{\tilde{h}}_{\mathrm{LoS}}\) is computed, through~\eqref{eq:h}. In the channel estimation stage, the CP-Mamba model takes the positioning result \(\mathbf{\tilde{C}}\) and the LoS channel \(\mathbf{\tilde{h}}_{\mathrm{LoS}}\) as priors to reconstruct the NLoS component \(\mathbf{\tilde{h}}_{\mathrm{NLoS}}\), thereby reducing pilot overhead and enhancing channel estimation performance.

\vspace{-5pt}
\subsection{Mamba Architecture}
To effectively handle near-field channel interference and model long-range dependencies in signals, we introduce the Mamba block based on the selective state space model (SSM)~\cite{Ma2024UMamba}. Building upon the classical continuous-time SSM, which converts a one-dimensional input sequence \(u(t)\) to an output \(y(t)\) via a linear ordinary differential equation, we have
\begin{equation}\label{eq:ssm_continuous}
	\begin{split}
		\frac{\mathrm{d}s(t)}{\mathrm{d}t}
		&= \mathbf{A}\,s(t) + \mathbf{B}\,u(t), \\[4pt]
		y(t)
		&= \mathbf{C}\,s(t),
	\end{split}
\end{equation}
where \(\mathbf{A}\), \(\mathbf{B}\), and \(\mathbf{C}\) denote the state, input, and output matrices, respectively, $s(t)$ denotes the implicit latent state.

\begin{figure}[!t]
	\centering
	\captionsetup{skip=2pt}
	\includegraphics[width=3.25in]{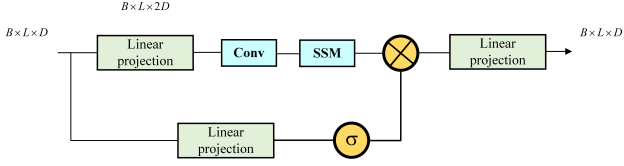}
	\caption{The architecture of the Mamba block.}
	\label{Fig4}
\end{figure}

Fig.~\ref{Fig4} illustrates the specific design of the Mamba block. Suppose the input tensor is $x \in \mathbb{R}^{B \times L \times D}$, where $B, L$ and $D$ represent the batch size, sequence length and feature dimension, respectively. The proposed model employs a dual-branch dynamic architecture for efficient long-range dependency modeling. First, $x$ is decomposed into parallel features $
[x_{b},\,z_{b}] \in \mathbb{R}^{B \times L \times 2D}$ via a linear projection. In the main branch, $x_{b}$ undergoes depthwise convolution (CONV) and sigmoid-weighted linear unit (SiLU) activation, followed by joint parameterization to generate time-varying parameters $
\delta \in \mathbb{R}^{B \times L \times d_{\mathrm{tr}}} $ and state matrices  \(\mathbf{B}\), \(\mathbf{C}\) $ \in \mathbb{R}^{B \times L \times d_{\mathrm{state}}}$. The temporal coefficient $\delta$ is discretized via a Softplus-mapped projection layer to produce
$\Delta$. \(\mathbf{A}\) selective scanning operator then integrates $x_{b},\;\Delta,\;$\(\mathbf{B}\),\;and\;\(\mathbf{C}\), aggregating multi-scale spatiotemporal features into $y_{s} \in \mathbb{R}^{B \times L \times 2D}$. In parallel, the auxiliary branch processes $
z_{b}$ using sigmoid gating to modulate $y_{s}$, yielding $y_{g} \in \mathbb{R}^{B \times L \times 2D}$. Finally, a linear projection restores the dimensionality to $y \in \mathbb{R}^{B \times L \times D}$.

\begin{figure}[!t]
	\centering
	\captionsetup{skip=-1pt}
	\includegraphics[width=3.5in]{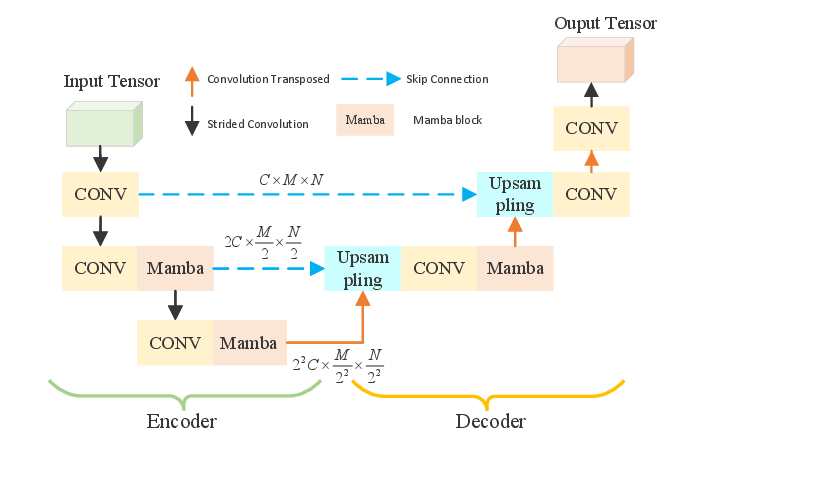}
	\caption{Proposed CP-Mamba network backbone.}
	\label{Fig5}
	\vspace{-10pt}
\end{figure}

\vspace{-5pt}
\subsection{CP-Mamba for positioning and channel estimation}
Building on the Mamba block, Fig.~\ref{Fig5} presents the proposed CP-Mamba network, which integrates the encoder-decoder and skip-connection principles of U-Net \cite{Ma2024UMamba} with the global-dependency modeling capability of Mamba. In the proposed CP-Mamba, the encoder integrates two-dimensional CONV layers and a Mamba layer. In the encoder network \(F_{\Theta_E}(\cdot)\), the first encoder block employs CONV layers to extract local features. For the remaining encoder blocks, CONV layers with stride 2 downsample spatial dimensions and increase feature channels, while the Mamba block focuses on capturing fine-grained local details and on improving resolution recovery. In the decoder network \(F_{\Theta_D}(\cdot)\), upsampling layers are applied to restore the spatial resolution of the feature maps. Finally, another CONV layer serves as the output layer of the decoder network \(F_{\Theta_D}(\cdot)\). The U-shaped architecture includes long skip connections between encoder and decoder blocks, enhancing feature reuse and preserving spatial details.

The positioning stage aims to estimate the UE's coordinate using pilot signals transmitted by the BS. In this stage, the encoder takes the pilot input $\mathbf{Y}_i$ and outputs a feature vector $\mathbf{\tilde{Y}}_i^{U}$, through~\eqref{eq:y}, which is then passed through the decoder followed by a pooling layer to predict the UE's coordinate $\mathbf{\tilde{C}}_i$. During the positioning stage, the system predominantly relies on the LoS path to determine the coordinates of the UE. Once the UE location is obtained, the LoS channel can be computed using geometric relationships. In contrast, during the channel estimation stage, both LoS and NLoS paths must be considered to accurately reconstruct the full channel.

In the channel estimation stage, the encoder takes the predicted UE's coordinates $\mathbf{\tilde{C}}_i$ and LoS channel $\mathbf{\tilde{h}}_i^{\mathrm{LoS}}$ as input and produces a feature vector $\mathbf{\tilde{u}}_i$. This vector is then passed to the decoder to reconstruct the NLoS component $\mathbf{\tilde{h}}_i^{\mathrm{NLoS}}$. By combining the reconstructed NLoS component with the LoS channel, the complete channel $\mathbf{\tilde{h}}_i$ is obtained. Meanwhile, we utilize the mean square error (MSE) metric as the loss function for learning the latent representation of the high-dimensional data. The MSE loss function between $\mathbf{h}_i$ and $\mathbf{\tilde{h}}_i$ can be expressed as
\begin{equation}
	C^P = \frac{1}{V} \sum_{i=1}^{V} \left\|\mathbf{h}_i  - \mathbf{\tilde{h}}_i \right\|^2
	\label{eq:cp-loss},
\end{equation}
where $V$ is the number of batch samples.

\section{Numerical Results}	
In our simulation, we set $N = 128$, $M = 64$, $V = 32$, $N_\mathrm{RF} = 4$, $P = 16$, $K=-0.00045 $, $f_c=28$\,GHz and $B=500$\,MHz. We assume that the UE is positioned around the BS such that $r\sim\mathcal U(0.1~\text{m},10~\text{m})$ and $\theta\sim\mathcal U(-90^{\circ},0^{\circ})$. We evaluate three array configurations CA, USA, NA and MOA with 20,000 training samples each. The NA uses a two level structure with 4 elements at level one and 124 at level two. The MOA consists of 16 modules of 8 elements each and  modular size $\Gamma=9$, balancing optimal spatial sampling and implementation feasibility. We adopt the mean positioning error (MPE) as the performance metric for positioning $\mathrm{MPE}=\mathbb{E}\!\left\{\lVert \hat{\mathbf{C}}_i-\mathbf{C}_i  \rVert_{F}\right\}$~\cite{Shi2024XL-MIMO}. The normalized MSE (NMSE) is used as the performance evaluation metric for channel estimation $\mathrm{NMSE} =\mathbb{E}\!\left\{ \frac{\lVert \hat{\mathbf{H}}_{k}-\mathbf{H}_{k} \rVert_{F}^{2}}{\lVert \mathbf{H}_{k} \rVert_{F}^{2}} \right\}$~\cite{UMLP_Xiao2023}. 

\begin{figure}[!t]
	\centering
	\captionsetup{skip=-1pt}
	\includegraphics[width=2.5in]{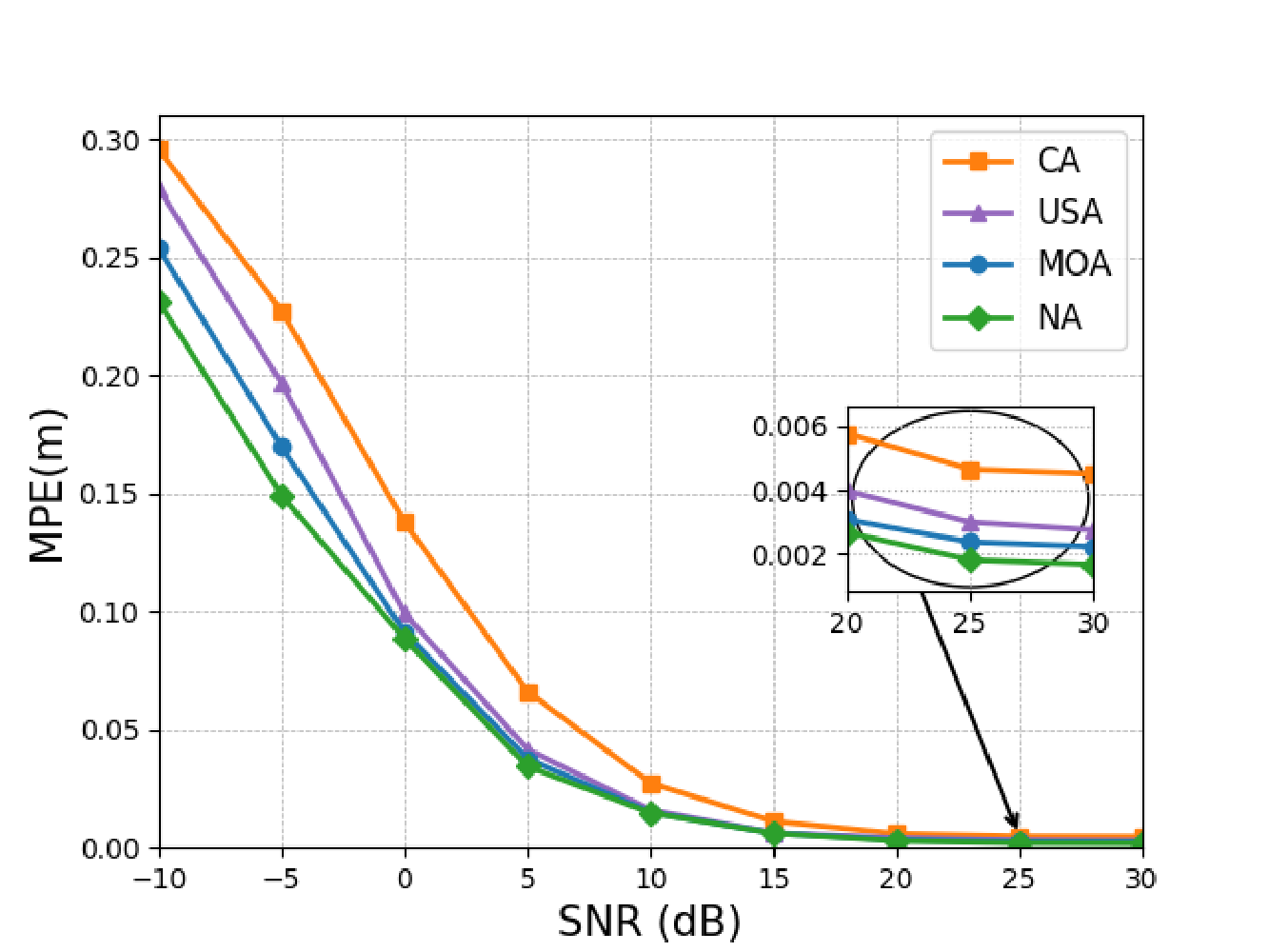}
	\caption{MPE performance for different antenna array of CP-Mamba.}
	\label{Fig6}
	\vspace{-10pt}
\end{figure}

\begin{figure}[!t]
	\centering
	\captionsetup{skip=-1pt}
	\includegraphics[width=2.5in]{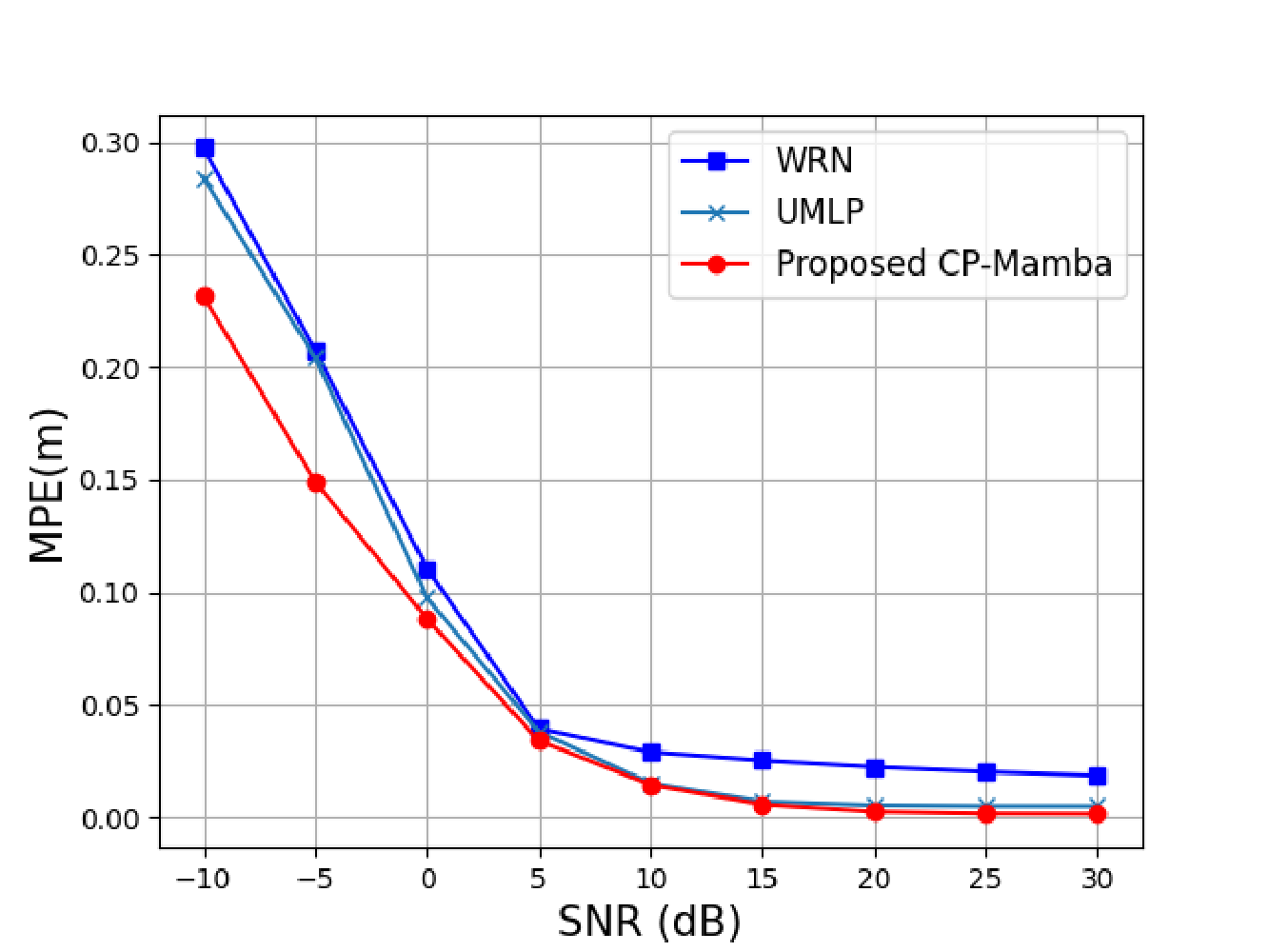}
	\caption{MPE performance for different methods with NA.}
	\label{Fig7}
	\vspace{-10pt}
\end{figure}

In Fig.~\ref{Fig6}, we evaluate the performance of the proposed CP-Mamba based method in LoS scenarios under CA and SAs. Under identical SNR conditions, SAs significantly surpass the CA in positioning accuracy, benefiting from their large apertures. Fig.~\ref{Fig7} compares the proposed CP-Mamba method with various benchmark schemes, including the DL-based WRN~\cite{Kim2020Transfer} and U-MLP~\cite{UMLP_Xiao2023} models, assuming an NA architecture. It is observed that CP-Mamba consistently achieves thr highest positioning accuracy across all SNR levels.
Additionally, Table~\ref{tab:addlabel} summarizes the number of parameters and floating-point operations (FLOPs) of different networks. Compared to other benchmarks, CP-Mamba has smaller parameters and less FLOPs, yet achieves superior channel estimation performance, demonstrating its efficiency and scalability.

\begin{figure}[!t]
	\centering
	\captionsetup{skip=-1pt}
\includegraphics[width=2.5in]{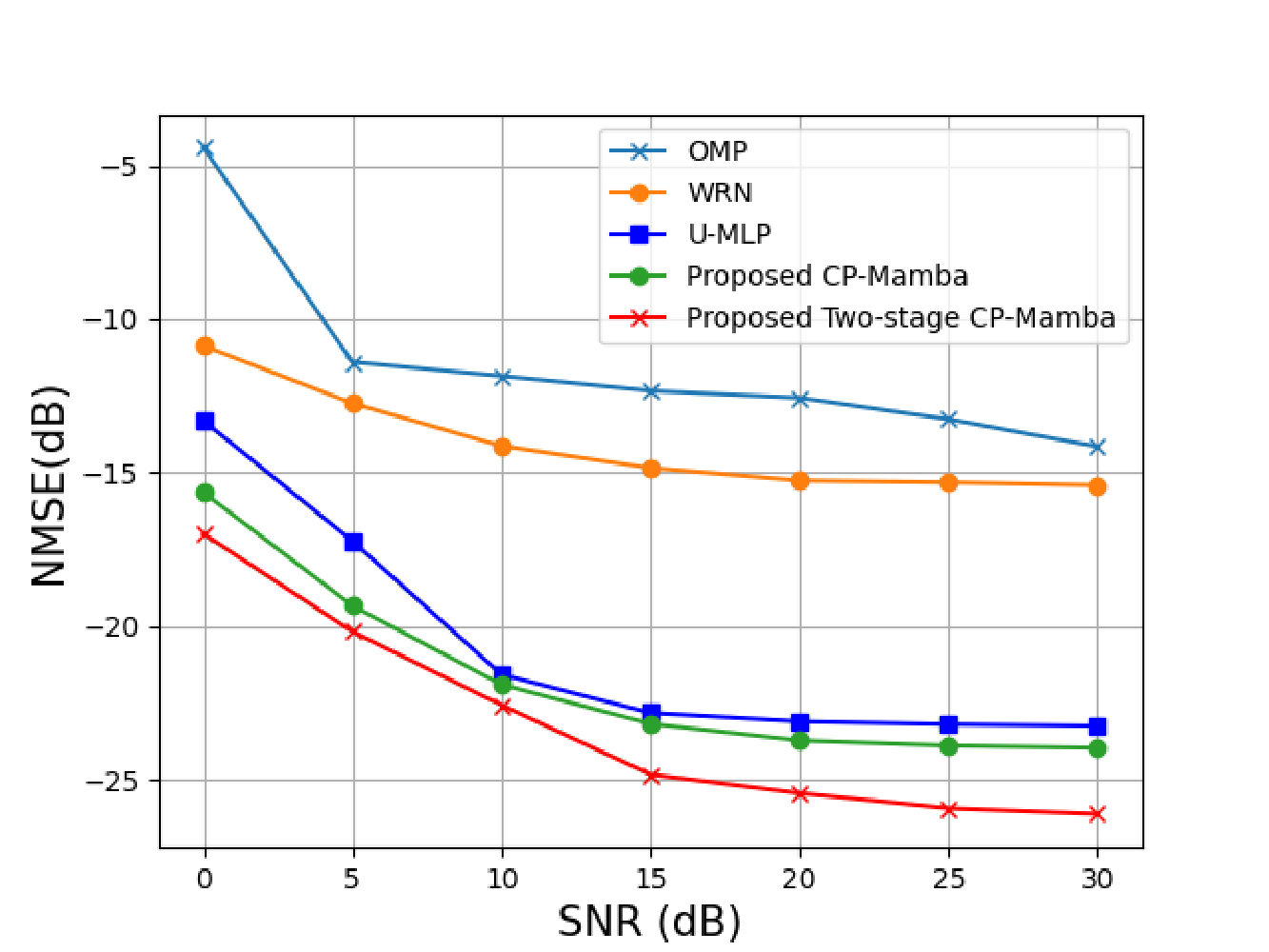}
	\caption{NMSE performance for different methods with NA.}
	\label{Fig8}
	\vspace{-10pt}
\end{figure}

Fig.~\ref{Fig8} compares the NMSE performance of the proposed CP-Mamba method with benchmark methods. The proposed method delivers superior channel estimation performance than the benchmarks across all SNR regimes. This improvement is attributed to the positioning stage of our scheme, which provides relatively accurate UE position estimates, a capability not offered by other methods, thereby enabling the integration of UE positional information to significantly enhance channel estimation accuracy.

\begin{table}[!t]
	\centering
	\caption{Floating-point Operations and Parameter Count.}
	\begin{tabular}{lcc}
		\toprule
		Algorithm & FLOPs (G) & Parameters (M) \\
		\midrule
		U-MLP[17]      & 3.60  & 3.07 \\
		WRN[18]    & 89.8 & 10.96 \\
		\textbf{CP-Mamba} & \textbf{1.96} & \textbf{1.71} \\
		\bottomrule
	\end{tabular}
	\label{tab:addlabel}
	\vspace{-10pt}
\end{table}


\section{Conclusion}

In this paper, we proposed a two-stage framework based on CP-Mamba for joint positioning and channel estimation. The first stage estimated UE coordinates, which were then used to enhance channel estimation in the second stage, enabling effective positioning-channel estimation synergy. To realize this, we designed the CP-Mamba network, which integrates U-Net for local feature extraction with Mamba blocks for modeling long-range dependencies. To evaluate the impact of SA antennas, three SA configurations were investigated. Numerical results showed that the proposed framework outperformed existing baselines, achieving higher positioning accuracy and channel estimation performance. Furthermore, SA antennas consistently outperformed CAs in both tasks.

\bibliographystyle{IEEEtran}
\bibliography{ZC}

\end{document}